\documentclass[aoas]{imsart}
\pdfoutput=1

\usepackage{amsthm,amsmath,natbib}
\usepackage{graphicx}
\usepackage{url}
\RequirePackage[colorlinks,citecolor=blue,urlcolor=blue]{hyperref}

\startlocaldefs

\newcommand{\kin}{k^{\rm in}}
\newcommand{\kout}{k^{\rm out}}
\newcommand{\gi}{{R_0}}
\newcommand{\eff}{{E_{\rm max}}}
\newcommand{\N}{{\rm N}}
\newcommand{\HN}{{\rm N^+}}

\newcommand{\Rss}{R^{\rm ss}}
\newcommand{\logit}{\mbox{logit}}
\newcommand{\invlogit}{\mbox{logit}^{-1}}

\endlocaldefs

\begin{document}

\begin{frontmatter}

\title{Bayesian aggregation of average data:\\An application in drug development}
\runtitle{Bayesian aggregation of average data}

\begin{aug}
\author{\fnms{Sebastian} \snm{Weber}\thanksref{m1}\corref{}\ead[label=e1]{sebastian.weber@novartis.com}}, %
\author{\fnms{Andrew} \snm{Gelman}\thanksref{m2,t1}\ead[label=e2]{gelman@stat.columbia.edu}}, %
\author{\fnms{Daniel} \snm{Lee}\thanksref{m3}\ead[label=e3]{daniel@generable.com}},\\
\author{\fnms{Michael} \snm{Betancourt}\thanksref{m2,t1}\ead[label=e4]{betanalpha@gmail.com}}, %
\author{\fnms{Aki} \snm{Vehtari}\thanksref{m4,t2}\ead[label=e5]{Aki.Vehtari@aalto.fi}}
\and
\author{\fnms{Amy} \snm{Racine-Poon}\thanksref{m1}\ead[label=e6]{amy.racine@novartis.com}}

\thankstext{t1}{Institute for Education Sciences R305D140059-16,
  Office of Naval Research N00014-15-1-2541 \& N00014-16-P-2039, Sloan
  Foundation G-2015-13987, National Science Foundation CNS-1205516,
  Defense Advanced Research Projects Agency DARPA BAA-16-32}

\thankstext{t2}{Academy of Finland (grant 298742)}

\affiliation{Novartis Pharma AG\thanksmark{m1},
  Columbia University\thanksmark{m2},\\
  Generable\thanksmark{m3} and
  Aalto University\thanksmark{m4}}

\address{Novartis Pharma AG\\
  Basel, Switzerland\\
  \printead{e1}\\
\phantom{E-mail:\ }\printead*{e6}}
\address{Department of Statistics\\
  Columbia University\\
  New York, USA\\
  \printead{e2}\\
\phantom{E-mail:\ }\printead*{e4}}
\address{Generable\\
  New York, USA\\
  \printead{e3}}
\address{Helsinki Institute for\\
  Information Technology HIIT\\
  Department of Computer Science\\
  Aalto University, Finland\\
  \printead{e5}}

\runauthor{Weber et al.}

\doi{10.1214/17-AOAS1122}

\end{aug}

\begin{abstract}

  Throughout the different phases of a drug development program,
  randomized trials are used to establish the tolerability, safety,
  and efficacy of a candidate drug. At each stage one aims to optimize
  the design of future studies by extrapolation from the available
  evidence at the time. This includes collected trial data and
  relevant external data. However, relevant external data are
  typically available as averages only, for example from trials on
  alternative treatments reported in the literature. Here we report on
  such an example from a drug development for wet age-related macular
  degeneration. This disease is the leading cause of severe vision
  loss in the elderly. While current treatment options are
  efficacious, they are also a substantial burden for the
  patient. Hence, new treatments are under development which need to
  be compared against existing treatments.

  The general statistical problem this leads to is {\em
    meta-analysis}, which addresses the question of how we can combine
  datasets collected under different conditions. Bayesian methods have
  long been used to achieve partial pooling. Here we consider the
  challenge when the model of interest is complex (hierarchical and
  nonlinear) and one dataset is given as raw data while the second
  dataset is given as averages only. In such a situation, common
  meta-analytic methods can only be applied when the model is
  sufficiently simple for analytic approaches. When the model is too
  complex, for example nonlinear, an analytic approach is not
  possible. We provide a Bayesian solution by using simulation to
  approximately reconstruct the likelihood of the external summary and
  allowing the parameters in the model to vary under the different
  conditions. We first evaluate our approach using fake-data
  simulations and then report results for the drug development program
  that motivated this research.

\end{abstract}



\end{frontmatter}


\section{Introduction}
Modern drug development proceeds in stages to establish the
tolerability, safety, and efficacy of a candidate drug
\citep{sheiner_learning_1997}. At each stage it is essential to plan
the next steps, using all relevant information. The collected raw data
are measurements of individual patients over time. Pharmacometric
models of such raw data commonly use nonlinear longitudinal
differential equations with hierarchical structure (also known as
population models), which can, for example, describe the response of
patients over time under different treatments. Such models typically
come with assumptions of model structure and variance components that
offer considerable flexibility and allow for meaningful extrapolation
to new trial designs. While these models can be fit to raw data, we
often wish to consider additional data which may be available only as
averages or aggregates. For example, published summary data of
alternative treatments are critical for planning comparative
trials. Such external data would allow for indirect comparisons as
described in the Cochrane Handbook \citep{higgins_cochrane_2011}.

Methods for the mixed case of individual patient data and aggregate
data are recognized as important, but are limited in their scope so
far. For example, in the field of pharmaco economics, treatments need
to be assessed which have never been compared in a head-to-head
trial. Methods such as Matching-Adjusted Indirect Comparisons (MAIC)
\citep{signorovitch_comparative_2010} and Simulated Treatment
Comparisons (STC) \citep{caro_no_2010,ishak_simulation_2015} have been
proposed to address the problem of mixed data in this domain. The
focus of these methods is a retrospective comparison of treatments
while we seek a prospective comparison under varying designs. That is,
in the MAIC approach the individual patient data is matched to the
reported aggregate data using baseline covariates. While simple in its
application, its utility is limited for a prospective planning of new
trials which vary in design properties. The STC approach offers
additional flexibility as it is based on the simulation of an index
trial to which other trials are matched using predictive
equations. However, the approach requires calibration for which
individual patient data is recommended. Hence, the effort of an STC
approach is considerable and its flexibility is still limited, since
the simulated quantities are densities of the endpoints. In contrast,
longitudinal nonlinear hierarchical pharmacometric models have the
ability to simulate the individual patient response over time and
hence give the greatest flexibility for prospective clinical trial
simulation, which provides valuable input to strategic decisions for a
drug development program.

Here we report on an example of a drug development program to
investigate new treatment options for wet age-related macular
degeneration (wetAMD), see
\citep{Ambati2012,Buschini2011,Khandhadia2012,Kinnunen2012}. This
disease is the leading cause of severe vision loss in the elderly
\citep{Augood2006}. Available drugs include anti-vascular endothelial
growth factor (anti-VEGF) agents which are repeatedly administered as
direct injections into the vitreous of the eye. The first anti-VEGF
agent was Ranibizumab
\citep{brown_ranibizumab_2006,rosenfeld_ranibizumab_2006}, with
another, Aflibercept \citep{heier_intravitreal_2012}, introduced
several years later. Initially, anti-VEGF intravitreal injections were
given monthly, and more flexible schemes with longer breaks between
dosings evolved over recent years to reduce the burden for patients
and their caregivers. In addition, a reduced dosing frequency also
increases compliance to treatment, which ensures sustained long-term
efficacy.

A key requirement for any new anti-VEGF agent is an optimized dosing
scheme to compare favorably to existing treatment options. For a
prospective evaluation of new trials, we simulate clinical trials
using nonlinear hierarchical pharmacometric models in which a new
anti-VEGF agent is compared to available treatments with various
design options. Important design options include the patient
population characteristics and the dosing regimen, which specifies
what dose amount is to be administered at which time-points to a given
patient.

In clinical studies, visual acuity is assessed by the number of
letters a patient can read from an ETDRS (Early Treatment Diabetic
Retinopathy Study) chart, expressed as best-corrected visual acuity
(BCVA) score, where the patient is allowed to use glasses for the
assessment. A nonlinear pharmacometric drug-disease model is able to
longitudinally regress the efficacy response as a function of the
patients' characteristics and individual dosing history. This
flexibility reduces confounding (through covariates and accounting for
non-compliance) during inference and enables realistic extrapolation
to future designs with alternative dosing regimens. However, these
models do require certain raw data that are commonly not reported in
the literature.  In our example, raw patient data from Ranibizumab
trials were available to us, but we only had aggregate data available
for Aflibercept. This creates the awkward situation that the reported
aggregate data on Aflibercept cannot be used to obtain accurate model
predictions despite our understanding that the nonlinear model is
appropriate for the same patient population and we are moreover only
interested in population predictions, i.e. the interest lies in
population parameters and not in patient specific parameters. The
problem is that the likelihood function for the aggregated data in
general has no closed-form expression. The standard
Expectation-Maximization or Bayesian approach in this case is to
consider the unavailable individual data points as missing data, but
this can be computationally prohibitive as it will vastly increase the
dimensionality of the problem space in an experiment with hundreds of
patients and multiple measurements per patient.

This paper describes how we enabled accurate clinical trial
simulations to inform the design of future studies in wetAMD, which
aim at improving the dosing regimens of anti-VEGF agents. This led us
to develop a novel statistical computational approach for integrating
averaged data from an external source into a linear or nonlinear
hierarchical Bayesian analysis. The key point is that we use an
approximate likelihood of the external average data instead of using
an approximate prior derived from the external data. Doing so enables
coherent joint Bayesian inference of raw and summary data.  The
approach takes account of possible differences in the model in the two
datasets.

In section \ref{methods} we describe the data and model for our study,
and section \ref{baad} lays out our novel approach for including
aggregate data into the pharmacometric model. Section \ref{simulation}
demonstrates our approach using simulation studies of a linear and a
nonlinear example. In the linear example we compare our approach to an
exact analytic reference, the nonlinear case is constructed to be
similar in its properties to the actual pharmacometric model. We
present results for our main problem in section \ref{pharm} and
conclude with a discussion in section \ref{discussion}.


\section{Data and Pharmacometric Model}\label{methods}

\subsection{Study data}\label{data}

We included in the analysis data set the raw data from the studies
MARINA, ANCHOR and EXCITE
\citep{rosenfeld_ranibizumab_2006,brown_ranibizumab_2006,schmidt-erfurth_efficacy_2011}. In
MARINA and ANCHOR a monthly Q4w treatment with Ranibizumab was
compared to placebo and active control, respectively. In MARINA a high
and a low-dose regimen treatment arm with Ranibizumab were included in
the trial. The EXCITE study tested the feasibility of an alternative
dosing regimen with longer Q12w (3 months) treatment intervals after
an initial 3 month loading phase of monthly treatments with
Ranibizumab. We restricted our analysis to the efficacy data only for
up to one year which is the follow-up time for the primary endpoints
of these studies. We consider the reported BCVA measure of the number
of letters read from the ETDRS chart which contains 0--100 letters.

\begin{figure}
\includegraphics{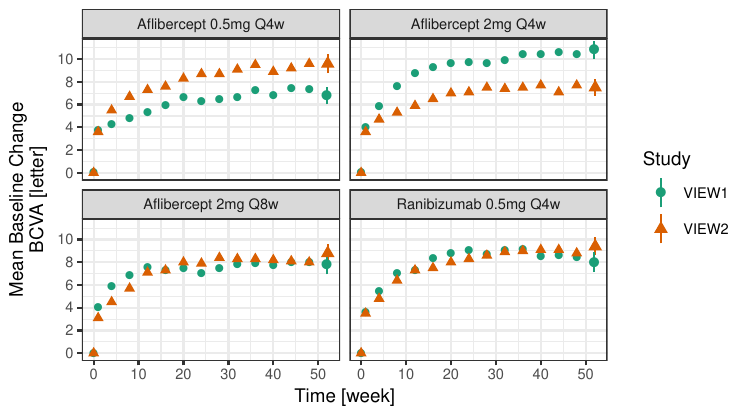}
\caption{Published average data of the VIEW1+2 studies
  (\cite{heier_intravitreal_2012}). Shown is the reported mean
  baseline change best-corrected visual acuity (BCVA) over a time
  period of one year. The vertical line at the last time point marks
  one standard error of the reported mean.}
  \label{fig:wetAMDraw}
\end{figure}

For Aflibercept no raw data from patients are available in the public
domain; only literature data of reported mean responses are available
from the VIEW1 and VIEW2 studies
\citep{heier_intravitreal_2012}. These studies assessed
non-inferiority of a low/high dose Q4w and a Q8w dosing regimen with
Aflibercept in comparison to 0.5mg Q4w Ranibizumab treatment,
which was also included in these studies as reference arm. Figure
\ref{fig:wetAMDraw} shows the reported mean BCVA data of VIEW1+2. In
table~\ref{tab:wetAMDdata} we list the baseline characteristics for
all the included study arms in the analysis.

\begin{table}
  
\begin{tabular}{lllrlrrr}
  Study & Data & Compound & N & Freq. & Dose & BCVA (SD) & Age (SD)\\
   &  &  &  &  & [mg] & [letter] & [y] \\
\hline
MARINA & patient & Ranibizumab & 238 & Q4w & 0.3 & 53.1 (12.9) & 77.4 (7.6)\\
MARINA & patient & Ranibizumab & 239 & Q4w & 0.5 & 53.7 (12.8) & 76.8 (7.6)\\
MARINA & patient & Placebo & 236 & Q4w & sham & 53.9 (13.7) & 77.1 (6.6)\\
\hline
ANCHOR & patient & Ranibizumab & 137 & Q4w & 0.3 & 47.1 (12.8) & 77.3 (7.3)\\
ANCHOR & patient & Ranibizumab & 139 & Q4w & 0.5 & 47.1 (13.2) & 75.9 (8.5)\\
\hline
EXCITE & patient & Ranibizumab & 120 & Q12w & 0.3 & 55.8 (11.8) & 75.1 (7.5)\\
EXCITE & patient & Ranibizumab & 118 & Q12w & 0.5 & 57.7 (13.1) & 75.8 (7.0)\\
EXCITE & patient & Ranibizumab & 115 & Q4w & 0.3 & 56.5 (12.2) & 75.0 (8.3)\\
\hline
VIEW1 & average & Aflibercept & 301 & Q4w & 0.5 & 55.6 (13.1) & 78.4 (8.1)\\
VIEW1 & average & Aflibercept & 304 & Q4w & 2.0 & 55.2 (13.2) & 77.7 (7.9)\\
VIEW1 & average & Aflibercept & 301 & Q8w & 2.0 & 55.7 (12.8) & 77.9 (8.4)\\
VIEW1 & average & Ranibizumab & 304 & Q4w & 0.5 & 54.0 (13.4) & 78.2 (7.6)\\
\hline
VIEW2 & average & Aflibercept & 296 & Q4w & 0.5 & 51.6 (14.2) & 74.6 (8.6)\\
VIEW2 & average & Aflibercept & 309 & Q4w & 2.0 & 52.8 (13.9) & 74.1 (8.5)\\
VIEW2 & average & Aflibercept & 306 & Q8w & 2.0 & 51.6 (13.9) & 73.8 (8.6)\\
VIEW2 & average & Ranibizumab & 291 & Q4w & 0.5 & 53.8 (13.5) & 73.0 (9.0)\\
\end{tabular}
  \caption{Baseline data of trials included in the analysis. The reported
    baseline BCVA and age are the respective mean values and their standard
    deviations.}
  \label{tab:wetAMDdata}
\end{table}

\subsection{Pharmacometric model}\label{model}

We use a drug-disease model which is informed on the basis of raw
measurements of individual patients over time. Such a model
\citep{weber_bayesian_2014} was developed on the available raw data
for Ranibizumab using the studies MARINA, ANCHOR and EXCITE.  The
visual acuity measure (BCVA) is limited to the range of 0--100
(letters read from the ETDRS chart) and so we modeled it on a
logit-transformed scale, $R_j(t) = \logit(y_{jk}/100)$, where $y_{jk}$
is the measurement for patient $j$ at time $t=x_k$.  The drug-disease
model used was derived from the semi-mechanistic turnover model
\citep{jusko_physiologic_1994}, which links a drug concentration,
$C_j(t)$, with a pharmacodynamic response, $R_j(t)$. The drug
concentration, $C_j(t)$, is determined by the dose amount and dosing
frequency as defined by the regimen. In our case the drug
concentration, $C_j(t)$, is latent, since no measurements of $C_j(t)$
in the eye of a patient is possible for ethical and practical
reasons. Therefore, we used a simple mono-exponential elimination
model and fixed the vitreous volume to $4$mL \citep{jr_adlers_1992}
and the elimination half-life $t_{1/2}$ from the vitreous to $9$ days
\citep{xu_pharmacokinetics_2013}. The standard turnover model assumes
that the response $R_j(t)$ can only take positive values, which is not
given on the logit-transformed scale. A modified turnover model is
therefore used, which is defined by the ordinary differential equation
(ODE)
\begin{equation}
\label{eq:turnMod}
\frac{dR_j(t)}{dt} =  \kin_j - \kout_j \left[ R_j(t) - \eff_j \, S_j(C_j(t))\right].
\end{equation}
The drug effect enters this equation via the function $S_j$, which is
typically chosen to be a Hill function of the concentration
$C_j(t)$. The Hill function is a logistic function of the $\log$ drug
concentration, $\invlogit(\log EC50 - \log C_j(t))$.  At baseline,
$R_j(t=0) = \gi_j$ defines the initial condition for the ODE. The
model in Eq. \eqref{eq:turnMod} has an important limit whenever a
time-constant stimulation, $S_j(t) = s_j$, is applied. Then, the ODE
system drives $R_j(t)$ towards its stable steady-state, which is
derived from Eq. \eqref{eq:turnMod} by setting the left-hand side to
$0$, $\Rss_j = (\kin_j/\kout_j) + \eff_j \, s_j$. In absence of a drug
treatment no stimulation is present; that is, $S_j(t) = s_j = 0$,
hence the ratio $\kin_j/\kout_j$ is of particular importance, as for
placebo patients it holds that
$\lim_{t\rightarrow \infty} R_j(t) = \kin_j/\kout_j$. The drug-disease
model describes treated patients in relation to placebo patients and
separates the drug-related parameters ($t_{1/2}$, $\eff$ and $EC50$)
from the remaining non-drug related parameters.

\section{Bayesian aggregation of average data}
\label{baad}

\subsection{General formulation}
\label{formulation} We shall work in a hierarchical
Bayesian framework. Suppose we have data
$y=(y_{jk}; j=1,\dots, J; k=1,\dots,T)$ on $J$ individuals at $T$ time
points, where each $y_j=(y_{j1},\dots,y_{jT})$ is a vector of data
with model $p(y_j|\alpha_j,\phi)$. Here, each $\alpha_j$ is a vector
of parameters for individual $j$, and $\phi$ is a vector of shared
parameters and hyperparameters, so that the joint prior is
$p(\alpha,\phi)=p(\phi)\prod_{j=1}^Jp(\alpha_j|\phi)$, and the primary
goal of the analysis is inference for the parameter vector $\phi$.

We assume that we can use an existing computer program such as Stan
\citep{stan2017} to draw simulations from the posterior distribution,
$p(\alpha,\phi|y)\propto p(\phi)\prod_{j=1}^Jp(\alpha_j|\phi)
\prod_{j=1}^Jp(y_j|\alpha_j,\phi)$.

We then want to update our inference using an {\em external dataset},
$y^{\prime}=(y^{\prime}_{jk}; j=1,\dots, J^{\prime};
k=1,\dots,T^{\prime})$,
on $J^{\prime}$ individuals at $T^{\prime}$ time points, assumed to be
generated under the model,
$p(y^{\prime}_j|\alpha^{\prime}_j,\phi^{\prime})$.  There are two
complications:
\begin{itemize}
\item The external data, $y^{\prime}$, are modeled using a process
  with parameters $\phi^{\prime}$ that are similar to but not
  identical to those of the original data. We shall express our model
  in terms of the difference between the two parameter vectors,
  $\delta=\phi^{\prime}-\phi$.  We assume the prior distribution
  factorizes as $p(\phi,\delta)=p(\phi)p(\delta)$.

  We assume that all the differences between the two studies, and the
  populations which they represent, are captured in $\delta$. One
  could think of $\phi$ and $\phi^{\prime}$ as two instances from a
  population of studies; if we were to combine data from several
  external trials it would make sense to include between-trial
  variation using an additional set of hyperparameters in the
  hierarchical model.

\item We do not measure $y^{\prime}$ directly; instead we observe the
  time series of averages,
  $\bar{y}^{\prime} = (\bar{y}^{\prime}_1, \dots,
  \bar{y}^{\prime}_T)$.
  And, because of nonlinearity in the data model, we cannot simply
  write the model for the external average data,
  $p(\bar{y}^{\prime}|\alpha^{\prime},\phi^{\prime})$, in closed form.
\end{itemize}
This is a problem of meta-analysis, for which there is a longstanding
concern when the different pieces of information to be combined come
from different sources or are reported in different ways \cite[see,
for example,][]{higgins_borrowing_1996,dominici_meta-analysis_1999}.

The two data issues listed above lead to corresponding statistical
difficulties:
\begin{itemize}
\item If the parameters $\phi^{\prime}$ of the external data were
  completely unrelated to the parameters of interest, $\phi$---that
  is, if we had a noninformative prior distribution on their
  difference, $\delta$---then there would be no gain from including
  the external data into the model, assuming the goal is to learn
  about $\phi$.

  Conversely, if the two parameter vectors were identical, so that
  $\delta\equiv 0$, then we could just pool the two datasets.  The
  difficulty arises because the information is partially shared, to an
  extent governed by the prior distribution on $\delta$.

\item Given that we see only averages of the external data, the
  conceptually simplest way to proceed would be to consider the
  individual measurements $y^{\prime}_{jk}$ as missing data, and to
  perform Bayesian inference jointly on all unknowns, obtaining draws
  from the posterior distribution,
  $p(\phi,\delta,\alpha,\alpha^{\prime}|y,\bar{y}^{\prime} )$. The
  difficulty here is computational: every missing data point adds to
  the dimensionality of the joint posterior distribution, and the
  missing data can be poorly identified from the model and the average
  data; weak data in a nonlinear model can lead to a
  poorly-regularized posterior distribution that is hard to sample
  from.
\end{itemize}

As noted, we resolve the first difficulty using an informative prior
distribution on $\delta$. Specifically, we consider in the following
that not all components of $\phi$, but only a few components, differ
between the datasets, such that the dimensionality of $\delta$ may be
smaller than that of $\phi$. This imposes that some components of
$\delta$ are exactly $0$.

We resolve the second difficulty via a normal approximation, taking
advantage of the fact that our observed data summaries are
averages. That is, as we cannot construct the patient specific
likelihood contribution for the external data set,
$\prod_{j=1}^{J^{\prime}}p(y^{\prime}_j|\alpha^{\prime}_j,\phi^{\prime})$,
directly, instead we approximate this term by a multivariate normal,
$\mbox{N}(\bar{y}^{\prime}|\tilde{M}_s,\frac{1}{J^{\prime}}\tilde{\Sigma}_s)$
to be introduced below.

\subsection{Inclusion of summary data into the likelihood}\label{idea}
Our basic idea is to approximate the probability model for the
external average data, $p(\bar{y}^{\prime} |\phi^{\prime})$, by a
multivariate normal with parameters depending on
$\bar{y}^{\prime}$. For a linear model this is the
analytically exact representation of the average data in the
likelihood. For nonlinear models the approximation is justified by
the central limit theorem if the summary is an average over many data
points. This corresponds in essence to a Laplace approximation to the
marginalization integral over the unobserved (latent) individuals in
the external data set $y^\prime$ as
$p(\bar{y}^{\prime} |\phi^{\prime}) = \int \!p(\bar{y}^{\prime}
|\alpha^{\prime},\phi^{\prime}) d\alpha^{\prime}$.

The existing model on $y$ is augmented by including a suitably chosen
prior on the parameter vector $\delta$ and the $\log$-likelihood
contribution implied by the external average data $\bar{y}^{\prime}$.
As such, the marginalization integral must be evaluated in each
iteration $s$ of the MCMC run. Evaluating the Laplace approximation
requires the mode and the Hessian at the mode of the integrand. Both
are unavailable in commonly used MCMC software, including Stan. To
overcome these computational issues, we instead use simulated plug-in
estimates.
In each iteration $s$ of the MCMC run we calculate the Laplace
approximation of the marginalization integral as follows:
\begin{enumerate}
\item Compute $\phi^{{\prime}}_s = \phi_s + \delta_s$.
\item Simulate parameters $\tilde{\alpha}_j$ and then data
  $\tilde{y}_{jk}, j=1,\dots, \tilde{J}, k=1,\dots,T^{\prime}$, for
  some number $\tilde{J}$ of hypothetical new individuals, drawn from
  the distribution $p(y^{{\prime}}|\phi^{{\prime}}_s)$, corresponding
  to the conditions under which the external data were collected
  (hence the use of the same number of time points $T^{\prime}$). The
  $\tilde{J}$ individuals do {\em not} correspond to the $J^{\prime}$
  individuals in the external dataset; rather, we simulate them only
  for the purpose of approximating the likelihood of the external average data,
  $\bar{y}^{\prime}$, under these conditions. The choice of $\tilde{J}$ must
  be sufficiently large, as is discussed below.
\item Compute the mean vector and the $T^{\prime} \times T^{\prime}$
  covariance matrix of the simulated data $\tilde{y}$.  Call these
  $\tilde{M}_s$ and $\tilde{\Sigma}_s$.
\item Divide the covariance matrix $\tilde{\Sigma}_s$ by $J^{\prime}$
  to get the simulation-estimated covariance matrix for
  $\bar{y}^{\prime}$, which is an average over $J^{\prime}$
  individuals whose data are modeled as independent conditional on the
  parameter vector $\phi^{\prime}$.
\item Approximate the marginalization integral over the individuals in
  the external $y^\prime$ data set with the probability density of the
  observed mean vector of the $T^{\prime}$ external data points using
  the multivariate normal distribution with mean $\tilde{M}_s$ and
  covariance matrix $\frac{1}{J^{\prime}}\tilde{\Sigma}_s$, which are
  the plug-in estimates for the mode and the Hessian at the mode of the
  Laplace approximation. The density
  $\mbox{N}(\bar{y}^{\prime}|\tilde{M}_s,\frac{1}{J^{\prime}}\tilde{\Sigma}_s)$
  then represents the information from the external mean data.
\end{enumerate}

\subsection{Computational issues: tuning and convergence}
\label{sec:tuning}
For the simulation of the $\tilde{J}$ hypothetical new individuals we
do need random numbers which are independent of the model. However, as
Bayesian inference results in a joint probability density, we cannot
simply declare an extra set of parameters in our model during an MCMC
run. That is, we can only control for the prior density of these
parameters, but not so for the posterior density which is generated by
the sampler. This is an issue, as by construction of Hamiltonian Monte
Carlo (HMC), as used in Stan, no random numbers can be drawn
independently from the model during sampling. However, our algorithm
does not require that the random numbers change from iteration to
iteration. Hence, we can simply draw a sufficient amount of random
numbers per chain and include these as data for a given chain. As
consequence, different chains may converge to different distributions
due to different initial sets of random numbers. However, with
increasing simulation size $\tilde{J}$ the simulations have a
decreasing variability in their estimates, as the standard error
scales with $\tilde{J}^{-1/2}$. Therefore, the tuning parameter
$\tilde{J}$ must be chosen sufficiently large to ensure convergence of
all chains to the same result. This occurs once the standard error is
decreased below the overall MC error. Whenever $\tilde{J}$ is chosen
too small, standard diagnostics like $\widehat{R}$
\citep{Gelman+etal+BDA3:2013} will indicate nonconvergence. We assess
this by running each odd chain with $\tilde{J}$ and each even chain
with $2\,\tilde{J}$ hypothetical new individuals (typically we run 4
parallel MCMC chains as this is free on a four-processor laptop or
desktop computer). The calculation of $\widehat{R}$ then considers
chains with different $\tilde{J}$, and so a too low $\tilde{J}$ will
immediately be detected, in which case the user can increase
$\tilde{J}$.

For models with a Gaussian residual error model, step 2 above can be
simplified. Instead of simulating observed fake data $\tilde{y}$, it
suffices to simulate the averages of the hypothetical new individuals
$\tilde{J}$ at the $T^\prime$ time-points. The residual error term can
be added to the variance-covariance matrix $\tilde{\Sigma}_s$ as
diagonal matrix. Should the sampling model not be normal, then normal
approximations should be considered to use. The benefit is a much
reduced simulation cost in each iteration of the MCMC run.


\section{Simulation Studies}\label{simulation}

\subsection{Hierarchical linear regression}\label{fake}

We begin with a fake-data hierarchical linear regression example which
is simple enough that we can compare our approximate inferences to a
closed form analytic solution to the problem as the unobserved raw
data can be marginalized over in a full analytic approach. We set up
this example to correspond in its properties to the longitudinal
nonlinear drug-disease model.

We consider a linear regression using a continuous covariate $x$
(corresponding to time) with an intercept, a linear, and a quadratic
slope term. The intercept and linear slope term vary in two ways which
is by individual and dataset. The quadratic term does not vary by
individual or dataset. This allows us to check two aspects: (a) if we
can learn differences between datasets (intercept and slope) and (b)
if the precision on fully shared parameters (quadratic term) increases
when combining datasets. That is, for the main dataset $y$, the model
is
$y_{jk}\sim\mbox{N}(\alpha_{j1} + \alpha_{j2} \, x_k + \beta \, x_k^2,
\sigma^2_y)$,
with prior distribution
$\alpha_{j} \sim\mbox{N}(\mu_{\alpha},\Sigma_{\alpha})$ for which we
set the correlations $\rho_{\alpha_{j1}\alpha_{j2}}$ (the off-diagonal
elements of $\Sigma_{\alpha}$) to $0$.  Using the notation from
section \ref{formulation}, the vector of shared parameters $\phi$ is
$\phi=(\mu_{\alpha1},\mu_{\alpha2},\beta,\sigma_{\alpha1},\sigma_{\alpha2},\sigma_y)$.
We assume that the number of individuals $J$ is large enough such that
we can assign a noninformative prior to $\phi$.

For the external dataset $y^{\prime}$, the model is
$y^{\prime}_{jk}\sim\mbox{N}(\alpha^{\prime}_{j1} + \alpha^{\prime}_{j2} \, x_k
+ \beta \, x_k^2, \sigma^2_y)$,
with the prior distribution
$\alpha^{\prime}_{j}\sim\mbox{N}(\mu_{\alpha}^{\prime},\Sigma_{\alpha})$.
In this simple example, we assign a noninformative prior distribution
to $\delta=\mu_\alpha^{\prime}-\mu_\alpha$ while we assign a $\delta$
of exactly $0$ to all other components in $\phi$ such that
$\phi'=(\mu_{\alpha1} +
\delta_{1},\mu_{\alpha2}+\delta_{2},\beta,\sigma_{\alpha1},\sigma_{\alpha2},\sigma_y)$.

\paragraph{Assumed parameter values}
We create simulations assuming the following conditions, which we set
to roughly correspond to the features of the drug-disease model:
\begin{itemize}
\item $J=100$ individuals in the original dataset, each measured $T=13$
  times (corresponding to measurements once per month for a year),
  $x_k=0,\frac{1}{12},\dots,1$.
\item $J^{\prime}=100$ individuals in the external dataset, also
  measured at these 13 time points.
\item $(\mu_{\alpha1},\sigma_{\alpha1})=(0.5, 0.1)$, corresponding to
  intercepts that are mostly between 0.4 and 0.6.  The data from our
  actual experiment roughly fell on a 100-point scale, which we are
  rescaling to 0--1 following the general principle in Bayesian
  analysis to put data and parameters on a unit scale
  \citep{gelman_parameterization_2004}.
\item $(\mu_{\alpha2},\sigma_{\alpha2})=(-0.2, 0.1)$, corresponding to
  an expected loss of between 10 and 30 points on the 100-point scale
  for most people during the year of the trial.
\item $\rho_{\alpha_{j1}\alpha_{j2}} = 0$,  no correlation 
  assumed between individual slopes and intercepts.
\item $\beta=-0.1$, corresponding to an accelerating decline
  representing an additional drop of 10 points over the one-year
  period.
\item $\sigma_y=0.05$, indicating a measurement and modeling error on
  any observation of about 5 points on the original scale of the data.
\end{itemize}
Finally, we set $\delta$ to $(0.1, 0.1)$, which represents a large
difference between the two dataset in the context of this problem, and
allows us to test how well the method works when the shift in
parameters needs to be discovered from data.

In our inferences, we assign independent unit normal priors for all
the parameters $\mu_{\alpha1}$, $\mu_{\alpha2}$, $\beta$, $\delta_1$, and $\delta_2$;
and independent half unit normal priors to the variance components
$\sigma_{\alpha1}$, $\sigma_{\alpha2}$, and $\sigma_y$. Given the
scale of the problem (so that parameters should typically be less than
1 in absolute value, although this is not a hard constraint), the unit
normals represent weak prior information which just serves to keep the
inferences within generally reasonable bounds.

\paragraph{Conditions of the simulations}
We run 4 chains using the default sampler in Stan, the HMC variant
No-U-Turn Sampler (NUTS)
\citep{hoffman_no-u-turn_2014,betancourt_diagnosing_2016}, and set
$\tilde{J}$ to 500 so that every odd chain will simulate 500 and every
even 1000 hypothetical individuals, thus allowing us to easily check
if the number of internal simulations is enough for stable
inference. If there were notable differences between the inferences
from even and odd chains, this would suggest that $\tilde{J}=500$ is
not enough and should be increased.

\paragraph{Computation and results} 
We simulate data $y$ and $y^{\prime}$. For simplicity we do our
computations just once in order to focus on our method only.  If we
wanted to evaluate the statistical properties of the examples, we
could nest all this in a larger simulation study.

\begin{figure}
\includegraphics{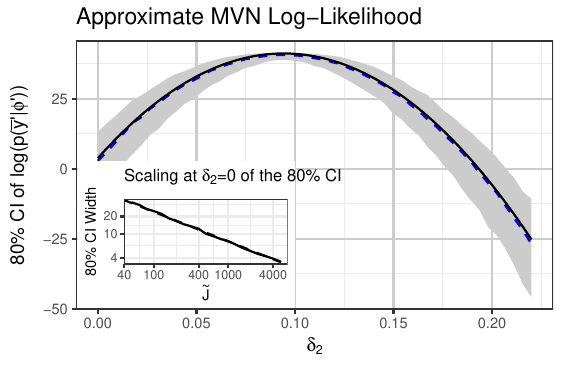}
\includegraphics{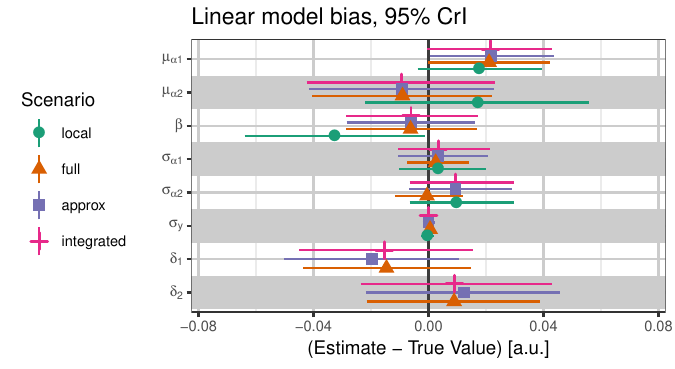}
\caption{Hierarchical linear model example. (Top) Comparison of the
  analytical expression for $\log p(\bar{y}^{\prime} |\phi^{\prime})$,
  shown as a solid black line, to the simulation based multivariate
  normal approximation
  $\mbox{N}(\bar{y}^{\prime}|\tilde{M}_s,\frac{1}{J^{\prime}}\tilde{\Sigma}_s)$. The
  simulation includes $\tilde{J} = 10^2$ hypothetical individuals, and
  $10^3$ replicates were performed to assess its distribution. The
  gray area marks the 80\% confidence interval and the dotted blue
  line is the median of the simulations. The inset shows the width of
  the 80\% confidence interval at $\delta_2=0$ as a function of the
  simulation size $\tilde{J}$ on a $\log$-$\log$ scale. The dotted
  line has a fixed slope of $-1/2$ and the intercept was estimated
  using least squares. (Bottom) The model estimates are shown as bias
  for the four different scenarios as discussed in the text. Lines
  show the 95\% credible intervals of the bias and the center point
  marks the median bias. The MCMC standard error of the mean is for
  all quantities below $10^{-3}$.}
\label{linear1}
\end{figure}

We first evaluate the simulation based approximation
of the log-likelihood contribution of the mean data
$\bar{y}^{\prime}$. This is shown in the top panel of Figure
\ref{linear1}. The plot shows $\log p(\bar{y}^\prime|\phi^\prime)$
evaluated at the true value of $\phi^\prime$ for varying values of
$\delta_2$. The gray band marks the 80\% confidence interval of $10^3$
replicates when simulating per replicate a randomly chosen set of
$\tilde{J} = 10^2$ patients. The dotted blue line is the median of
these simulations and the black solid line is the analytically
computed expression for $\log p(\bar{y}^\prime|\phi^\prime)$ which we
can compute for this simple model directly. Both lines match
respectively which suggests that the simulation approximation is
consistent with the analytical result. The width of the gray band is
determined by the number of hypothetical fake patients
$\tilde{J}$. The inset plot shows at a fixed value of $\delta_2 = 0$
the width of the 80\% confidence interval as a function of $\tilde{J}$
in a $\log$-$\log$ plot. The solid black line marks the simulation
results while the dashed line has a fixed slope of $-1/2$ and a
least-squares estimated intercept. As both lines match each other, we
can conclude that the scaling of the confidence interval width is
consistent with $\propto \tilde{J}^{-1/2}$.

We run the algorithm as described below and reach approximate
convergence in that the diagnostic $\widehat{R}$ is near
$1$ for all the parameters in the model. We then compare the
inferences for the different scenarios:
\begin{description}
\item[local:] The posterior estimates for the shared parameters $\phi$
  using just the model fit to the local data $y$.
\item[full:] The estimates for all the parameters $\phi,\delta$ using
  the complete data $y,y^{\prime}$, which would not in general be
  available---from the statement of the problem we see only the
  averages for the new data $y^{\prime}$---but we can do so here as we
  have simulated data.
\item[approx:] The estimates when using the approximation scheme for
  all the parameters $\phi,\delta$ using the actual available data
  $y,\bar{y}^{\prime}$.
\item[integrated:] The estimates when using an analytical likelihood
  for all the parameters $\phi,\delta$ using the actual available data
  $y,\bar{y}^{\prime}$. In general it would not be possible to compute
  this inference directly, as we need the probability density for the
  averaged data, but in this linear model this distribution has a
  closed-form expression which we can calculate.
\end{description}
The bottom panel of Figure \ref{linear1} shows the results of the
parameter estimates as bias. We are using informative priors and so we
neither desire nor expect expect a bias of exactly $0$. Rather we
would like to see for each parameter a match of the approximate
estimate ({\em blue line with a square}) with the estimate of the full
scenario ({\em orange line with a triangle}), which corresponds to the
correct Bayes estimate.  However, we cannot expect that the full
scenario matches the approximate estimate, since the correct Bayes
estimate for the full scenario is given by
$p(\phi,\delta|y,y^{\prime})$ which is based on the individual raw
data $y$ and $y^\prime$ instead of $y$ and mean data
$\bar{y}^\prime$. The appropriate comparison is wrt to the integrated
scenario ({\em red line with a cross}) which is the correct Bayes
estimate of $p(\phi,\delta|y,\bar{y}^{\prime})$. The integrated and
the approximate scenarios do match closely for all parameters.

When comparing the full scenario with the approximate and integrated
result one can observe that the variance components
$\sigma_{\alpha 1}$ and $\sigma_{\alpha 2}$ are estimated with higher
precision in the full scenario. This is a direct consequence of using
the reported means only for the external data.

Including the averaged data $\bar{y}^{\prime}$ into the model does not
inform the variance components $\sigma_{\alpha 1}$ and
$\sigma_{\alpha 2}$, but it does increase the precision of all other
parameters in $\phi$. This can be observed by considering the reduced
width of the credible intervals when comparing the local scenario
({\em green line with a dot}) to the others, in particular for
$\mu_{\alpha 2}$ and $\beta$. The estimates of $\delta_1$ and
$\delta_2$ are similar across all cases whenever these can be
estimated. This suggests that the external averaged data
$\bar{y}^\prime$ are just as informative for the $\delta$ vector as
the individual raw data $y^\prime$ themselves. The main reason as to
why the precision of the $\delta$ estimate is a little higher for the
full scenario is related to the estimates of the variance components
$\sigma_{\alpha 1}$ and $\sigma_{\alpha 2}$. These variance components
are estimated from the complete individual raw data ($y$ and
$y^\prime$) to be smaller in comparison to the other scenarios. As a
result the overall weight of each patient to the log-likelihood is
larger. This leads to a higher precision of the population parameters
which can be observed in particular for the parameters
$\mu_{\alpha 1}$ and $\delta$.

\subsection{Hierarchical nonlinear pharmacometric model}
\label{pharmSim}

Next we perform a fake-data study that is closely adapted to our
application of interest. The function $R_j(t)$ in
Eq. \eqref{eq:turnMod} is only implicitly defined; no closed-form
solution is available for the general case. For the simulation study
we consider the special case of constant maximal drug effect at all
times; that is, $S_j(t) = s_j = 1$ for a patient $j$ who receives
treatment or $S_j(t) = s_j = 0$ for placebo patients otherwise. The
advantage of this choice is that the ODE can then be solved
analytically as
$ R_j(t) = \Rss_j + \left(\gi_j - \Rss_j\right) \exp{(- \kout_j t)}$.
In the following we consider 3 different cohorts of patients (placebo,
treatment 1 and 2) observed at times $t=x_k$. Data for treatment 2 will
be considered as the external dataset and given as average data only
to evaluate our approach. Measurements $y_{jk}$ of a patient $j$ are
assumed to be iid normal,
$y_{jk}/100 \sim \mbox{N}(\invlogit(R_{j}(x_k)), \sigma_y^2)$.  We
assume that the number of patients is large enough such that
weakly-informative priors, which identify the scale of the parameters,
are sufficient. The above quantities are parametrized and assigned the
simulated true values and priors for inference as:

\begin{itemize}
\item $J = 100$ patients in the data-set with raw measurements per
  individual patient. The first $j=1, \ldots, 50$ patients are assigned
  a placebo treatment ($\eff_j = 0$) and the remaining
  $j=51,\ldots, 100$ patients are assigned a treatment with nonzero
  drug effect ($\eff_j > 0$). All patients are measured at $T=13$ time
  points corresponding to one measurement per month over a year. We
  rescale time accordingly to $x_k=0, \frac{1}{12}, \ldots, 1$.
\item $J^\prime = 100$ patients in the external dataset, measured at the
  same $T^\prime=13$ time points.
\item $\gi_{j} \sim \N(L\alpha_0, \sigma_{L\alpha_0}^2)$ is the
  unobserved baseline value of each patient $j$ on the logit scale
  which we set to $L\alpha_0 = 0$ corresponding to $50$ on the
  original scale and $\sigma_{L\alpha_0} = 0.2$. We set the
  weakly-informative prior to $L\alpha_0 \sim \N(0, 2^2)$ and
  $\sigma_{L\alpha_0} \sim \HN(0,1^2)$.
\item $\kin_j/\kout_j = L\alpha_s$ is the placebo steady state, the
  asymptotic value patients reach if not on treatment (or treatment is
  stopped). In the example lower values of the response correspond to
  worse outcome. We set the simulated values to
  $L\alpha_s = \logit(35/100)$ and the prior to
  $L\alpha_s \sim \N(-1, 2^2)$.
\item $\log(1/\kout_j) \sim \N(l\kappa, \sigma_{l\kappa}^2) $
  determines the patient-specific time scale of the exponential
  changes ($\kout_j$ is a rate of change).  We assume that changes in
  the response happen within $10/52$ time units which led us to set
  $l\kappa = \log(10/52)$ and we defined as a prior
  $l\kappa \sim \N(\log(1/4), \log(2)^2)$ and
  $\sigma_{l\kappa} \sim \HN(0,1^2)$.
\item $\log(\eff_j)$ is the drug effect for patient $j$. If patient
  $j$ is in the placebo group, then $\eff_j = 0$. For patients
  receiving the treatment 1 drug we assumed
  $\log(\eff_j) = l\eff_j = \log(\logit(60/100) - \logit(35/100))$
  which represents a gain of $25$ points in comparison to
  placebo. Patients in the external data set $y^\prime$ are assumed to
  have received the treatment 2 drug and are assigned a different
  $l\eff^\prime$. We consider $\delta = l\eff^\prime - l\eff = 0.1$,
  which corresponds to a moderate to large difference
  ($\exp(0.1) \approx 1.1$).  As priors we use
  $l\eff \sim \N(\log(0.5), \log(2)^2)$ and $\delta \sim \N(0, 1^2)$.
\item $\sigma_y = 0.05$ is the residual measurement error on the
  original letter scale divided by $100$. The prior is assumed to be
  $\sigma_y \sim \HN(0, 1^2)$.
\end{itemize}

\begin{figure*}
\includegraphics{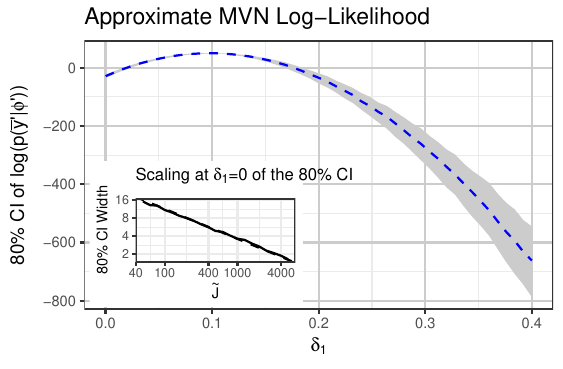}
\includegraphics{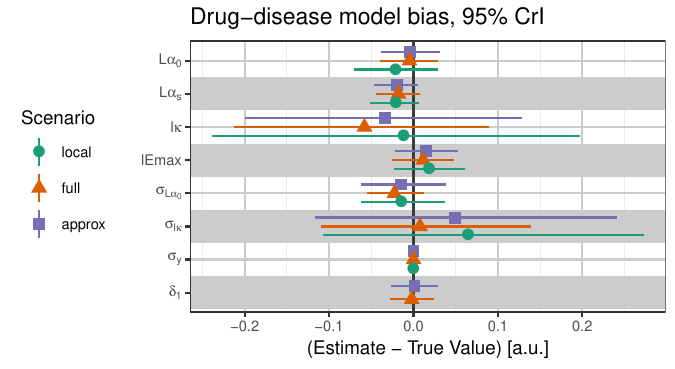}
\caption{Drug-disease model example: (Top) Assessment of the
  distribution of the multi-variate normal approximation to
  $\log p(\bar{y}^{\prime} |\phi^{\prime})$ at a simulation size of
  $\tilde{J} = 10^2$ hypothetical fake patients using $10^3$
  replicates for varying $\delta_1$. The gray area marks the 80\%
  confidence interval, the blue dotted line is the median of the
  simulations. The inset shows the width of the 80\% confidence
  interval at $\delta_1=0$ as a function of the simulation size
  $\tilde{J}$ on a $\log$-$\log$ scale. The dotted line has a fixed
  slope of $-1/2$ and the intercept was estimated using least
  squares. (Bottom) The model estimates are shown as bias for the three
  different scenarios as discussed in the text. Lines show the 95\%
  credible intervals of the bias and the center point marks the median
  bias. The MCMC standard error of the mean is for all quantities
  below $10^{-3}$.}
  \label{fig:estimatesPharm}
\end{figure*}

All simulation results are shown in Figure
\ref{fig:estimatesPharm}. In the top panel of
Fig. \ref{fig:estimatesPharm} an assessment of the sampling
distribution of our approximation is shown for a simulation size of
$\tilde{J} = 10^2$ hypothetical fake patients and $10^3$
replicates. Since for this nonlinear example we cannot integrate out
analytically the missing data in the external data set such that there
is no black reference line as before. However, we can conclude that
the qualitative behavior of a maximum around the simulated true value
is like in the linear case. Moreover, the inset confirms that the
scaling of the precision of the approximation with increasing
simulation size $\tilde{J}$ of hypothetical fake patients scales as a
power law consistent with $\propto \tilde{J}^{-1/2}$.

For the model we run 4 chains and set $\tilde{J}$ to $500$ as
before. The model estimates are shown as bias in the bottom panel of
Figure \ref{fig:estimatesPharm}. The precision of the estimates from
the local fit ({\em green line with a dot}) increases when adding the
external data. While population mean parameters gain in precision in
the full ({\em orange line with a triangle}) and approx ({\em blue
  line with a square}) scenario, the precision of variance component
parameters like $\sigma_{L\alpha_0}$ and $\sigma_{l\kappa}$ only
increase in the full scenario. This is expected as the mean data
$\bar{y}^\prime$ does not convey information on between-subject
variation. However, it is remarkable that the population mean
parameter estimates for the approx scenario are almost identical to
the full scenario, including the important parameter $\delta_1$.

We can conclude that possible differences in a drug-related parameter,
$\delta_1$, can equally be identified from individual raw data as from
the external mean data only. The mean estimate for $\delta_1$ and its
95\% credible interval in the full scenario $(y,y^\prime)$ and the
approximate scenario $(y,\bar{y}^\prime)$ do match one another
closely.

\section{Results for the Drug Development Application}\label{pharm}

We now turn to the application of our approach for the development of
a new drug for wetAMD. For Aflibercept no raw data from patients is
available in the public domain; only literature data of reported mean
responses are available \citep{heier_intravitreal_2012}. Hence,
extrapolation for Aflibercept treatments on the basis of the developed
drug-disease model was not possible. The drug related parameters of
the drug-disease model are the elimination half-life $t_{1/2}$, the
maximal drug effect, $lEmax$, and the concentration at which 50\% of
the drug effect is reached, $lEC50$ (both parameters are estimated on
the $\log$ scale). The elimination half-life is fixed with a drug
specific value in our model from values reported in the literature for
each drug. We can inform the latter two parameters for Ranibizumab
from our raw data which comprise a total of $N=1342$ patients from
the studies MARINA, EXCITE and ANCHOR; the data from the VIEW1+2
studies (\cite{heier_intravitreal_2012}, $N=1210+1202$) enables us to
estimate these parameters for Aflibercept. Following our approach, we
modified the existing model on Ranibizumab to include a $\delta$
parameter (with a weakly-informative prior of $\N(0,1)$) for each of
the drug related parameters for patients on Aflibercept treatment. In
addition, we also allowed the baseline BCVA of VIEW1+2 to differ as
compared to the chosen reference study MARINA. We did not include a
$\delta$ parameter for any other parameter in the model, since the
remaining parameters characterize the natural disease progression in
absence of any drug.  We consider it reasonable to assume that the
natural disease progression does not change under the two conditions,
and in any case it is impossible to infer differences in the natural
disease progression as compared to our dataset with the VIEW1+2 data
since no placebo patients were included in either study for ethical
reasons.

It is important to note that the VIEW1+2 studies included each a
0.5mg Q4w treatment arm with Ranibizumab. For these arms only the mean
data is reported as well and we include these into our model as a
reference, assuming that the drug-specifc parameters are exactly the
same for all datasets.

Figure \ref{fig:wetAMDraw} shows the published mean baseline change
BCVA data of the VIEW1+2 studies. From the VIEW1+2 studies we choose
to include only the mean BCVA data of the dosing regimens 2mg Q8w
Aflibercept and 0.5mg Q4w Ranibizumab into our model as these are used
in clinical practice and are hence of greatest interest to describe
these as accurately as possible. The total dataset then included raw
data from $N=1342$ patients from MARINA, ANCHOR and EXCITE (different
Ranibizumab regimens and a placebo arm) and $N=1202$ patients from the
reported mean data in VIEW1+2 (2mg Q8w Aflibercept and 0.5mg Q4w
Ranibizumab). Since our model is formulated on the scale of the
nominally observed BCVA measurements, we shifted the reported baseline
change BCVA values by the per study mean baseline BCVA value. We used
the remaining data from the 2mg Q4w and 0.5mg Q4w Aflibercept regimens
for an out-of-sample model qualification.

\begin{figure}
$\vcenter{\hbox{\includegraphics{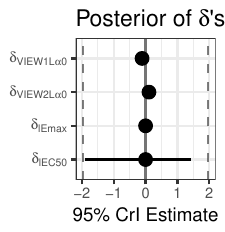}}}$%
$\vcenter{\hbox{\includegraphics{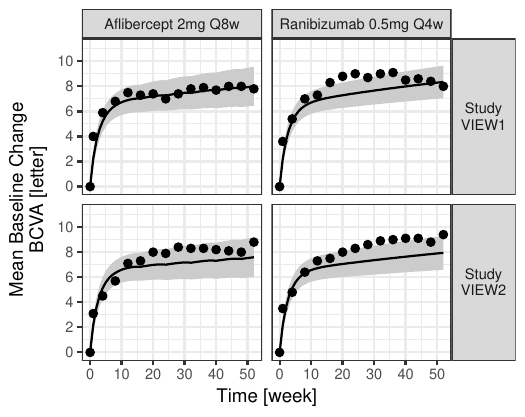}}}$
\caption{Main analysis results: (Left) Shows the posterior 95\%
  credible intervals of the estimated $\delta$ parameters. The dotted
  lines mark the 95\% credible interval of the prior. (Right) Shows
  the predicted mean baseline change BCVA as solid line for the study
  arms included in the model fit. The gray area marks one standard
  error for the predicted mean, assuming a sample size as reported per
  arm (about 300 each, see table~\ref{tab:wetAMDdata}). The dots mark
  the reported mean baseline change BCVA and are shown as reference.}
  \label{fig:wetAMD}
\end{figure}

The final result of the fitted model, which uses our internal
patient-level data and the VIEW1+2 summary data of the 2mg Q8w
Aflibercept and 0.5mg Q4w Ranibizumab arms, are shown in Figure
\ref{fig:wetAMD}. Presented are the posteriors of the $\delta$
parameters (left) and the posterior predictive of the mean baseline
change BCVA response of the two included regimens of VIEW1+2 (right).

The posterior predictive distribution of the mean baseline change BCVA
is in excellent agreement with the reported data for the 2mg Q8w
Aflibercept arms of VIEW1+2. The posterior predictive distribution of
the 0.5mg Q4w Ranibizumab mean data in VIEW1+2 suggests a slight
under-prediction from the model. However, the prediction is for one
standard error corresponding to a 68\% credible interval and hence the
observed data is well in the usual 95\% credible interval.

When comparing the posteriors of the $\delta$ parameters to their
standard-normal priors (corresponding to a prior 95\% credible
interval from $-1.96$ to $+1.96$), we observe that the information
implied by the aggregate data of VIEW1+2 for each parameter varies
substantially. While the $\delta_{lEmax}$ parameter is estimated with
great precision to be close to $0$, the precision of the
$\delta_{lEC50}$ posterior is only increased slightly from a prior
standard deviation of $1$ to a posterior standard deviation of
$0.8$. This is a consequence of the dosing regimens in VIEW1+2, which
keep patients at drug concentrations well above the $lEC50$ in order
to ensure maximal drug effect at all times. In fact, the only trial
in our Ranibizumab database where concentrations vary around the range
of the $lEC50$ is the EXCITE study. This study included two Q12w
Ranibizumab arms which showed a decrease of the BCVA after the loading
phase such that drug concentrations have apparently fallen below the
$lEC50$ which makes its estimation possible, see
\citep{schmidt-erfurth_efficacy_2011}.

\begin{figure}
\includegraphics{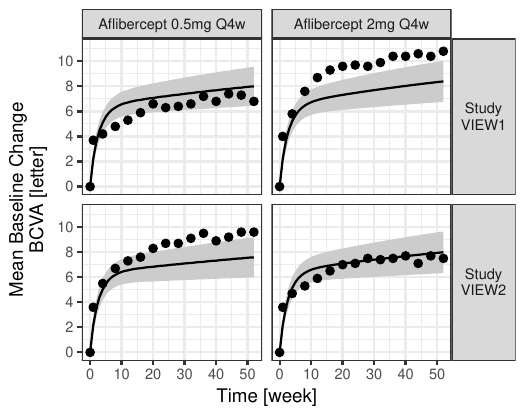}
\caption{Out-of-sample model qualification: Shown is the predicted
  mean baseline change BCVA as solid line for the study arms of
  VIEW1+2 which were not included in the model fitting. The gray area
  marks one standard error for the predicted mean assuming a sample
  size as reported per arm (about 300 each, see
  table~\ref{tab:wetAMDdata}). The dots mark the reported mean
  baseline change BCVA and are shown as reference.}
  \label{fig:wetAMDqual}
\end{figure}

The out-of-sample model qualifications are shown in
Figure~\ref{fig:wetAMDqual}. The 2mg Q4w Aflibercept of VIEW2 arm is
well predicted by the model, while the respective regimen in VIEW1 is
predicted less successfully. This arm was reported to have an
unusually high mean baseline change BCVA outcome for reasons which are
still not well understood such that we did not investigate
further. Moreover, the regimen 0.5mg Q4w Aflibercept appears to be
under-predicted in VIEW2 and slighlty over-predicted in
VIEW1. However, when considering that VIEW1+2 are exactly replicated
trials, the observed differences in this arm (see
Figure~\ref{fig:wetAMD}) are not expected (also note that the ordering
for each regimen reversed when comparing these in VIEW1 and VIEW2). If
we were to compare our model predictions against an averaged result
from VIEW1+2, these comparisons would look more favorable as the study
differences would average out. We conclude that the average outcomes
are well captured while the per arm variations are within limits which
are known and still unexplained.

In summary, our final model is able to predict accurately the 2mg Q8w
Aflibercept regimen which is our main focus when including the VIEW1+2
data into our analysis. The 2mg Q8w Aflibercept regimen is one of
the treatments for wetAMD applied in clinical practice.

\section{Discussion}
\label{discussion}

Model-based drug development hinges on the amount of information which
we can include into models. While hierarchical patient-level nonlinear
models offer the greatest flexibility, they make raw patient-level
data a requirement. This can limit the utility of such models
considerably, as relevant information may only be available to the
analyst in aggregate form from the literature. For our wetAMD drug
development program the presented approach enabled patient-level
clinical trial simulations for most wetAMD treatments used in the
clinic. Our approach was used to plan confirmatory trials which test a
new treatment regimen with less frequent dosing patterns against
currently established regimens. In particular, these results were used
to plan the confirmatory studies \cite{harrier_study} and
\cite{hawk_study} which evaluate Brolucizumab in comparison to
Aflibercept. These trials test a new and never observed dosing regimen
aiming at a reduced dosing frequency while maintaining maximal
efficacy. Within this regimen patients are assessed for their
individual treatment needs during a Q12w-learning cycle. Depending on
this assessment, patients are allocated to a Q12w or a Q8w schedule. A
key outcome of the trials is the proportion of patients allocated to
the Q12w regimen. Through the use of our approach it was possible to
include highly relevant information from the literature into a
predictive model which supported strategic decision making for the
drug development program in wetAMD.

The critical step in our analysis was to model jointly our study
data and external aggregate data. We constructed a novel Bayesian
aggregation of average data which had to overcome three
different issues:
\begin{enumerate}
\item Our new data were in aggregated average form; the raw data
  $y^{\prime}$ were not available, and we could not directly
  write or compute the likelihood for the observed average data
  $\bar{y}^{\prime}$.
\item The new data were conducted under different experimental
  conditions.  This is a standard problem in statistics and can be
  handled using hierarchical modeling, but here the number of
  ``groups'' is only 2 (the old data and the new data), so it would
  not be possible to simply fit a hierarchical model, estimating
  group-level variation from data.
\item It was already possible to fit the model to the original data
  $y$, hence it made sense to construct a computational procedure that
  made use of this existing fit.
\end{enumerate}

We handled the first issue using the central limit theorem (CLT),
which was justified by the large sample size of the external
data. This allowed us to approximate the sampling distribution of the
average data by a multivariate normal and using simulation to compute
the mean and covariance of this distribution, for any specified values
of the model parameters.

We handled the second issue by introducing a parameter $\delta$
governing the difference between the two experimental conditions. In
some settings it would make sense to assign a weak prior on $\delta$
and essentially allow the data to estimate the parameters separately
for the two experiments; in other cases a strong prior on $\delta$
would express the assumption that the underlying parameters do not
differ much between groups. Seen from a different perspective, the new
experimental condition is considered as a biased observation of an
already observed experimental condition, which goes back to
\citet{pocock_combination_1976}.

Finally, we formulated our approach by extending an existing
model. That is, we added a term to the log-likelihood of the original
model. This term represents the information from the external
means. We used a nested simulation scheme which we ran during the MCMC
fit. The key step to perform the nested simulation scheme was to
generate a sufficiently large sample of random numbers prior to the
MCMC run and to then use this sample for each iteration of the running
MCMC to perform effectively a Monte Carlo integration. We expect this
nested integration approach to be useful in general, since its
applicability is not restricted to the presented application of
marginalizing the likelihood over a latent variable space, but can be
applied in general during a MCMC run.

Our proposed approach is an approximate solution wrt to the
alternative approach, which is to represent the patient-level data of
the external dataset as latent. As our simulation studies have
revealed, we are still able to obtain accurate estimates of the
$\delta$ parameter vector, which is our main objective here. The
reason is the large sample size of the external data, which ensures
that the assumption of the CLT holds well. The use of our approximate
procedure does lead to a reduction of computational resources needed
to integrate the external average data. Thus, we can then use these
freed up computational resource to model more accurately the
patient-level data and obtain in return better predictions. As
external datasets of interest are usually of considerable sample size,
we expect this to be an advantageous choice to spend our finite
computational resources in these applications.

Considering our idea more generally, we have effectively reversed the
common Bayesian approach in which external data are commonly used to
elicit a prior which is then updated with experimental data through
the model likelihood. In our approach this paradigm is conceptually
reversed: the external data is explicitly made part of the model
likelihood which then informs our parameters of interest. In this
light, we expect that our ideas will allow for future developments of
general interest, such as the formulation of implicit priors or the
definition of an effective sample size for complex models using a
normal approximation.

In this work we have expanded the applicability of Bayesian
meta-analysis to the broad class of nonlinear hierarchical models for
the case whenever we wish to learn from aggregated average data, which
renders data from individuals latent and only indirectly reported via
means. This situation oftentimes arises in the domain of biostatistics
which uses meta-analytic approaches in various stages of drug
development. However, the ideas presented are general and should also
find application in other domains. For our specific case this work
enabled accurate clinical trial simulations which supported the design
of large phase III trials aiming to establish better treatments in
wetAMD.


\bibliography{all}

\appendix

\section*{Acknowledgements}
We thank Bob Carpenter for fruitful discussions on the
manuscript.

\begin{supplement}[id=suppA]
  \sname{Supplement A}
  \stitle{Program sources}
  \slink[doi]{10.1214/17-AOAS1122SUPP}
  \sdatatype{.zip}
  \sdescription{Source code of R and Stan programs of simulation
    studies and drug-disease model.}
\end{supplement}

\end{document}